\documentclass[a4paper,10pt]{article}

\usepackage{graphicx}
\usepackage{amsfonts}
\usepackage{stmaryrd}

\newtheorem{thm}{Theorem}
\newtheorem{lem}{Lemma}
\newtheorem{tab}{Table}
\newtheorem{fig}{Figure}

\newtheorem{cor}{Corollary}

\def\leurre{\noindent\leftskip0pt\small\baselineskip 10pt}

\def\encadre#1#2{%
\setbox100=\hbox{\kern#1{#2}\kern#1}
\dimen100=\ht100 \advance \dimen100 by #1
\dimen101=\dp100 \advance \dimen101 by #1
\setbox100=\hbox{\vrule height \dimen100 depth \dimen101\box100\vrule}
\setbox100=\vbox{\hrule\box100\hrule}
\advance \dimen100 by .4pt \ht100=\dimen100
\advance \dimen101 by .4pt \dp100=\dimen101
\box100
\relax
}

\def\ligne#1{\hbox to \hsize{#1}}
\def\PlacerEn#1 #2 #3 {\rlap{\kern#1\raise#2\hbox{#3}}}

\font\rmx=cmr10
\font\rmxii=cmr12

\font\itx=cmti10
\font\ttx=cmtt10

\def\qed{\hfill$\boxempty$}

\title{About embedded quarters and points at infinity in the hyperbolic plane
\vskip 15pt
\rmxii
\ligne{\hfill Maurice Margenstern\hfill} 
\vskip 15pt
\rmx\baselineskip=12pt
\ligne{\hfill
Laboratoire d'Informatique Th\'eorique et Appliqu\'ee, EA 3097,\hfill}
\ligne{\hfill Universit\'e de Lorraine,\hfill}
\ligne{\hfill Campus du Saulcy,\hfill}
\ligne{\hfill 57045 Metz Cedex, France,\hfill}
\ligne{\hfill {\itx email:} {\ttx maurice.margenstern@univ-lorraine.fr}\hfill}
}

\begin{document}
\maketitle

\vskip 10pt
\begin{abstract}
In this paper, we prove two results. First, there is a family of sequences of embedded 
quarters of the hyperbolic plane such that any sequence converges to a limit which is an
end of the hyperbolic plane. Second, there is no algorithm which would allow us to check
whether two given ends are equal or not.
\end{abstract}
\vskip 10pt

\def\cqfd{\hbox{\kern 2pt\vrule height 6pt depth 2pt width 8pt\kern 1pt}}
\def\Hii{$I\!\!H^2$}
\def\Hiii{$I\!\!H^3$}
\def\Hiv{$I\!\!H^4$}
\def\norm{\hbox{$\vert\vert$}}
\section{Introduction}

   This study takes place in hyperbolic geometry, in a specific tiling of the
hyperbolic plane, the tessellation $\{5,4\}$ which I called the \textbf{pentagrid},
see~\cite{mmbook1}.

   Fix such a tessellation. Denote by~$a$ the length of a side of a tile of the tessellation.
In this tiling, we call \textbf{quarter}, a subset of the tiling which is the intersection
of two half-planes whose lines support consecutive edges of a pentagon~$P$ of the tessellation.
This pentagon is called the \textbf{head} of the quarter and the common point of the lines
delimiting the half-planes is called the \textbf{vertex} of the quarter. Note that the
quarter is delimited by two rays issued from the vertex and supported by the above mentioned
lines. These rays are also the \textbf{border} of the quarter.

   In this paper we are interested by sequences of quarters such that each term of the sequence
is included in the next one. We shall show that the vertices of such quarters tend to
a limit. 
To this aim, Section~\ref{convinf} fixes the notion of neighbourhood for a point at infinity. 
Section~\ref{prelim}
studies simple properties of included quarters. But before, we had to establish
specific projection properties of the pentagrid in Section~\ref{cornucopia}.
Section~\ref{sequences} proves that
a sequence of embedded quarters has a limit and 
Section~\ref{nocomput} shows two results of undecidability concerning points at infinity.

\section{Prolegomenon: the cornucopia representation}
\label{cornucopia}

We fix~$O$ a point of the hyperbolic plane and two orthogonal rays issued from~$O$: $p$ and~$q$. We may assume that, counter-clockwise turning around~$O$, $p$ comes
before~$q$. We say that $p$ is \textbf{horizontal} and that~$q$ is \textbf{vertical}.
The rays~$p$ and~$q$ constitute the border of a \textbf{quarter} of the plane, $\cal Q$.
Such a quarter can also be viewed as the intersection of two half-planes whose borders
are perpendicular.

First, let us fix notations.
Consider a pentagon~$P$. Counter-clockwise and consecutively number the sides of~$P$ 
by $i$ with \hbox{$i\in[1..5]$}. Denote by~$\ell_i$ be the line which supports the 
side~$i$ and let $A$, $B$, $C$, $D$ and~$E$ be the vertices of~$P$, counter-clockwise 
labelled in this way, with $A$, $E$ belonging to both sides~5 and~1, sides~5 and~4
respectively.  Each line~$\ell_i$ defines two half-planes~$H_i$ and~$\neg H_i$. 
Let~$H_i$ denote the half-plane which contains~$P$. Call \textbf{lower strip} of~$P$ the region which is defined by~\hbox{$H_1\cap H_4\cap \neg H_5$}. In the lemmas of the 
paper, we shall speak of the side~$i$ of a pentagon, having in mind a numbering as the 
one we already considered for~$P$, and we shall always remind which side is side~1 in 
order to avoid ambiguities. Note that sides~1 and~4 are opposite and that~$\ell_5$ is the 
common perpendicular of~$\ell_1$ and~$\ell_4$.
\vskip 10pt
\vtop{
\ligne{\hfill
\includegraphics[scale=0.5]{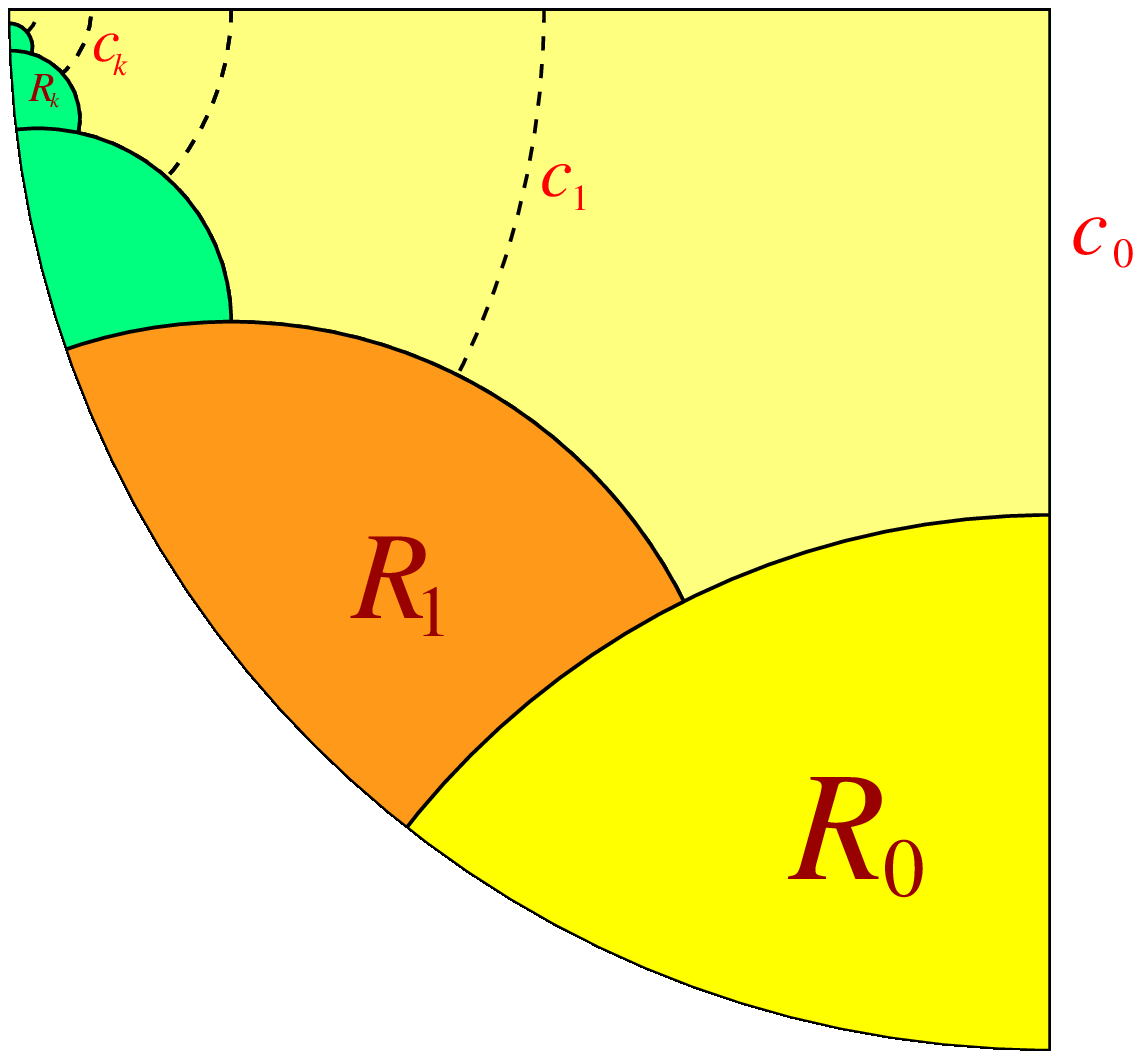}
\hfill
\includegraphics[scale=0.5125]{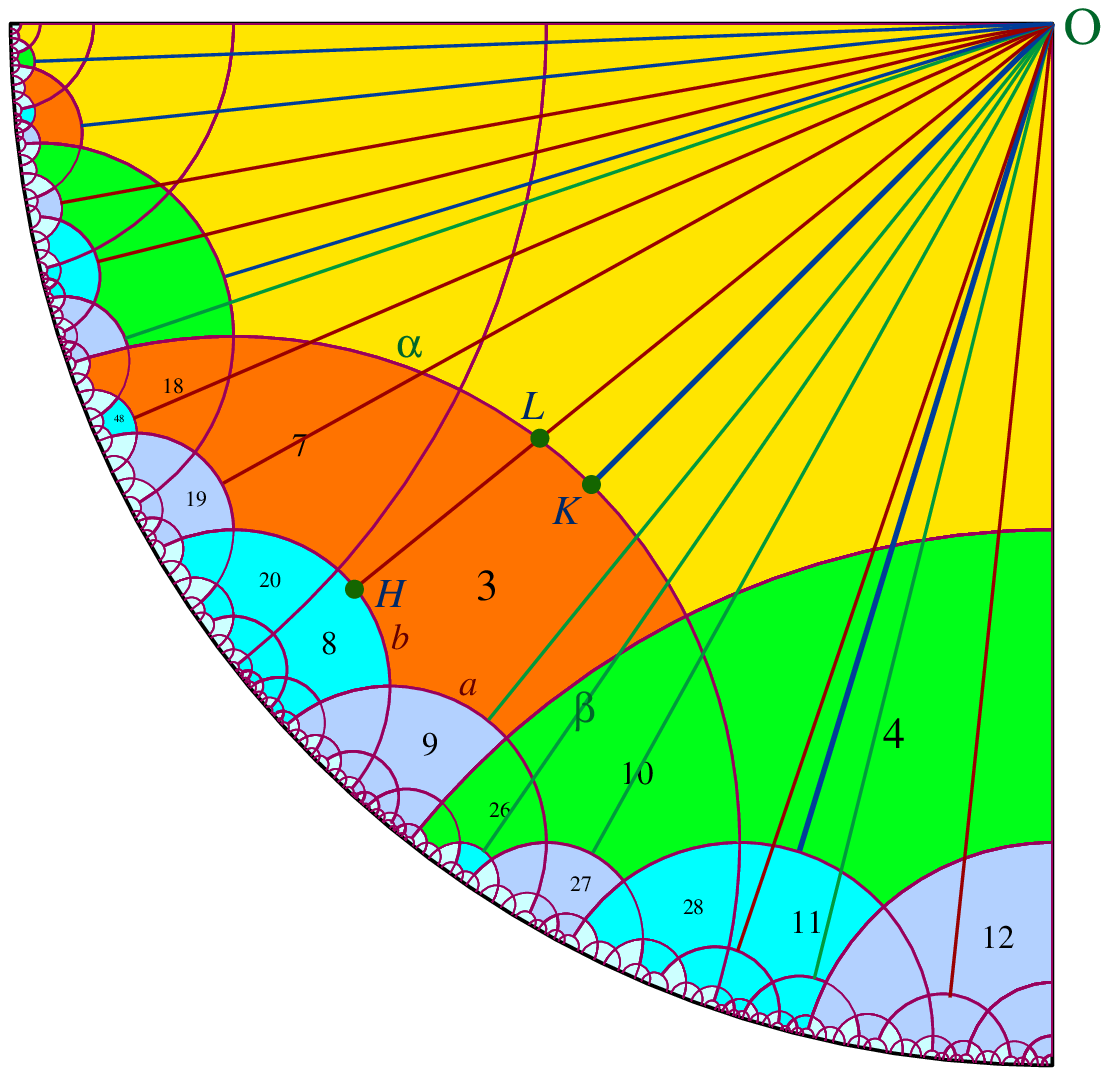}
\hfill}
\vspace{-5pt}
\begin{fig}
\label{bij}\leurre
To left: the cornucopia. 
\vskip 0pt
\noindent
To right: proving properties of the cornucopia.
\end{fig}
}

Let \hbox{\{$P_n\}_{n\in\mathbb{N}}$} denote a sequence of pentagons lying 
in~$\cal Q$ such that any $P_n$ has an edge contained in~$p$, such that~$P_0$ 
has~$O$ as a vertex and that its edges meeting at~$O$ are contained in~$p$ and~$q$, 
see Figure~\ref{bij}, and such that for any~$n$, $P_n$ and~$P_{n+1}$ have a common 
side: it is both the side~1 of~$P_n$ and the side~4 of~$P_{n+1}$. The complement
of the $P_n$'s in~$\cal Q$ can be represented as a union of quarters $R_n$ as 
illustrated by the left-hand side picture of~Figure~\ref{bij}. We call \textbf{cornucopia 
of~$\cal Q$} the union of the $P_n$'s.  The $R_n$'s are defined as follows: $R_0$ is 
bordered by~$q$ and by the line which support the side~3 of~$P_0$, $R_{n+1}$ is 
bordered by the line which supports the side~3 
of~$P_n$ and by the line  which supports the side~2 of~$P_n$. Remember that in a 
pentagon, sides~2 and~3 are perpendicular at the point where they meet.

The quarters $R_n$ can be defined in another way: $R_0$ is the image of~$\cal Q$ by 
the shift~$\tau_0$ along~$q$ of amplitude~$a$. Note that~$\tau_0$ transforms~$O$ in 
the other vertex of the side~4 of~$P_0$ which lies in~$q$. Note that the shift~$\tau$ 
along~$p$ of amplitude~$a$ transforms the side~1 of~$P_n$ into the side~1 
of~$P_{n+1}$. Now,  let ${\cal Q}_{n+1}$ be the image of ${\cal Q}_n$ by~$\tau$, 
putting \hbox{${\cal Q}_0=\cal Q$}. Then, $R_{n+1}$ is the image of~$R_n$ under~$\tau$. Note that  $R_{n+1}$ is also the image of ${\cal Q}_n$ by the shift~$\tau_n$ along the side~1 of~$P_n$ of amplitude~$a$. Note that $\tau_n$ 
translates this decomposition of~$\cal Q$ into each quarter~$R_{n+1}$. Consider the 
recursive iteration of this decomposition in all new quarters generated in this way. We say 
that the regions~$R_m$ belong to the first generation, so that the shift of the 
decomposition of~${\cal Q}_m$ in each of them by~$\tau_m$ defines the second generation. In a similar way, the generation $n$+1 is obtained from the 
generation~$n$. The decomposition of each region into the cornucopia and its 
complement constitute the \textbf{cornucopia decomposition} of~$\cal Q$. 

Presently, we wish to give a better algorithmic representation of the cornucopia decomposition of~$\cal Q$ which will allow us to prove interesting properties.

\begin{lem}\label{visilow}
Let~$R_2$ and~$R_3$
be the pentagons obtained from~$P$ by refection in its sides~$2$ and~$3$ respectively.
Define the side~$5$ of~$R_2$, $R_3$ to be the side~$2$, $3$ of~$P$ respectively.
Then the lower strip of~$R_2$, $R_3$ respectively, contains the lower strip of~$P$.
\end{lem}

\noindent
Proof.
Remember that the lower strip~$\cal S$ of~$P$ is defined as \hbox{$H_1\cap H_4\cap\neg H_5$}.
Note that $R_2$, $R_3$ is also the shift~$\tau_1$,$\tau_4$ respectively  of~$P$ along the 
side~1, 4, respectively, of~$P$ of amplitude~$a$, see Figure~\ref{lower_strip} where $P_0$
plays the role of~$P$.
Denote by~${\cal S}_2$, ${\cal S}_3$ the strip of~$R_2$, $R_3$ respectively.
Then, \hbox{${\cal S}_i=H_1^{\tau_{j_i}}\cap H_4^{\tau_{j_i}}\cap\neg{H_5}^{\tau_{j_i}}$}, with 
\hbox{$i\in\{2,3\}$}, \hbox{$j_2=1$} and \hbox{$j_3=4$}. We have that
\hbox{$H_1^{\tau_1}=H_1$}.  Now, $\tau_1$ can be decomposed into the reflection~$\beta$
in the bisector of side~1 followed by the reflection~$\rho$ in side~2. Now, $\beta$
transforms~$\ell_4$ into~$\ell_3$ and $\rho$ leaves~$\ell_3$ globally invariant, so that
\hbox{$H_4^{\tau_1}=H_3$}. We have too that $\beta$ transforms~$\ell_5$ into~$\ell_2$
and $\ell_2$ is invariant under~$\rho$. Consequently,
\hbox{$(\neg H_5)^{\tau_1}=H_2$}. 
Now, a product of two reflections in axes which are perpendicular to~$\ell_1$ shows that
\hbox{${\cal S}\subset H_3\cap H_2$}. 
Hence,
\hbox{${\cal S}\subset{\cal S}_2$}. Similarly, \hbox{$H_4^{\tau_4}=H_4$},
\hbox{$H_1^{\tau_4}=H_2$} and \hbox{$(\neg H_5)^{\tau_4}=H_3$}, so that we obtain that
\hbox{${\cal S}\subset{\cal S}_3$}.\qed

\vskip 5pt
\vtop{
\ligne{\hfill
\includegraphics[scale=0.7]{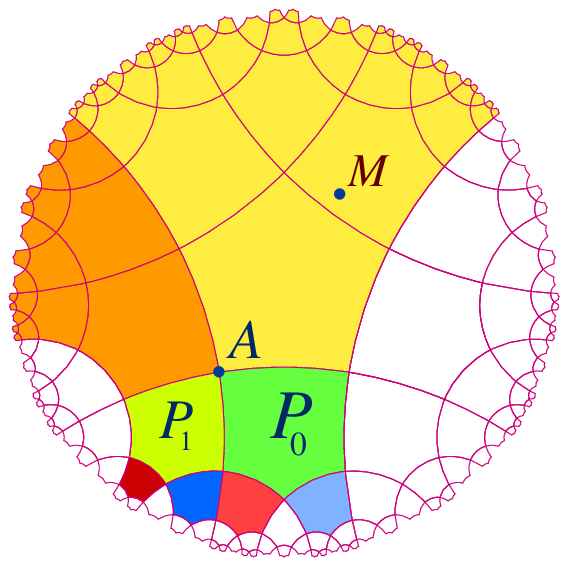}
\hfill}
\vspace{-5pt}
\begin{fig}
\label{lower_strip}\leurre
Illustration of the proofs of Lemma~{\rm\ref{visilow} and Lemma~{\rm\ref{visi_shift}}}. 
\end{fig}
}

\begin{lem}\label{visi_shift}
Consider the pentagon~$P$. Let~$\tau$ be the shift along the side~$5$ of~$P$ of amplitude~$a$,
transforming~$\ell_4$ into~$\ell_1$. Let~$Q$ be the image of~$P$ under~$\tau$. 
Let $S_2$, $S_3$ be the image of~$Q$ by reflection in its sides~$2$, $3$ respectively.
Let the side~$5$ of~$S_2$, $S_3$ be the side~$2$, $3$ of~$Q$ respectively. Then
the lower strip of~$S_2$ contains that of~$P$, but the strip of~$S_3$ does not meet
that of~$P$, except on the line~$\ell_4$ of~$P$.
\end{lem}
 
\noindent
Proof. Denote by $\sigma_1$, $\sigma_4$ the shift of amplitude~$a$ along the line~1, 4
of~$Q$ respectively, which transforms the side~5 of~$Q$ into its side~2, 3 respectively.
Denote by $i_Q$ the side~$i$ of~$Q$.
Denote by ${\cal T}_2$, ${\cal T}_3$ the lower strip of~$S_2$, $S_3$ respectively.
Repeating the proof of Lemma~\ref{visilow}, we obtain that
\hbox{${\cal T}_i=H_{1_Q}^{\sigma_{j_i}}\cap H_{4_Q}^{\sigma_{j_i}}
\cap(\neg H_{5_Q})^{\sigma_{j_i}}$}, with \hbox{$i\in\{2,3\}$}
and \hbox{$j_2=1$} and \hbox{$j_3=4$}. Repeating the same argument,
we get that \hbox{${\cal T}_2=H_{1_Q}\cap H_{3_Q}\cap (\neg H_{2_Q})$}.
Now, \hbox{$H_1\subset H_{1_Q}$} as \hbox{$H_{1_Q}=H_1^\tau$}. This is obtained by
decomposing~$\tau$ into the reflection~$\gamma$ in the bisector of side~5
followed by a the reflection~$\rho_1$ in the side~1.
similarly, we have that \hbox{$H_4\subset H_4^\tau=H_{4_Q}$}. At last, note 
that \hbox{$H_{5_Q}=H_5$}, so that
\hbox{$(\neg H_{5_Q})^{\sigma_1}=\neg H_{2_Q}$} and 
\hbox{$\neg H_5\subset \neg H_{2_Q}$}. From this we get that 
\hbox{${\cal S}\subset{\cal T}_2$}.
For~$S_3$, note that $H_{4_Q}\cap H_1\subset\ell_1$. Now, \hbox{$\ell_{4_Q}=\ell_1$}.
Accordingly, 
as \hbox{${\cal T}_3\subset H_{4_Q}=\neg H_1$}, ${\cal T}_3$ cannot meet $\cal S$.\qed
 
\begin{lem}\label{projinstrip}
Consider a pentagon~$P$ with its sides, their support and its vertices labelled as above
indicated. Let~$M$ be a point in~$\cal S$, the lower strip of~$P$. Let~$K$ be the 
orthogonal projection of~$M$ on~$\ell_5$. Then $K$ is in side~$5$. If $M$~is 
in~$\ell_1$ or in~$\ell_4$, then \hbox{$M=A$} or \hbox{$M=E$} respectively. 
\end{lem}

\noindent
Proof. Let $H_1$, $H_4$ be the half-plane defined by~$\ell_1$, $\ell_4$ respectively 
which contains~$P$. If \hbox{$K\not\in {\cal S}$}, then, by construction of~$H_1$ 
and~$H_4$ we have \hbox{$K\not\in H_1$} or \hbox{$K\not\in H_4$}. Assume that \hbox{$K\not\in H_1$}.
Then, $PK$ cuts~$\ell_4$ in~$T$. Whether \hbox{$T=E$} or \hbox{$T\not=E$}, from~$T$ we have two distinct perpendiculars to~$\ell_5$ which is impossible. 
A similar argument proves that $K$~cannot be in~$H_1$.\qed

\begin{lem}\label{projpenta}
Let~$P$ be a pentagon with the same labelling as in Lemma~{\rm\ref{projinstrip}}. Let~$M$ belong to the lower strip of~$P$. Let~$K$, $F$ and~$G$ be the orthogonal projection of~$M$ on~$\ell_5$, $\ell_2$ and~$\ell_3$ respectively.  
Then, $F$ belongs to side~$2$, $G$ belongs to side~$3$, $MF$ cuts~$\ell_5$ in the open segment $]AK[$ and~$MG$ cuts~$\ell_5$ in the open segment $]KE[$.
Note that if $M$ belongs to~$\ell_1$,$\ell_4$, then $K$ and~$F$,$G$ respectively
also belong to~$\ell_1$,$\ell_4$ respectively, and the conclusion for~$G$,$F$ 
respectively still holds.
\end{lem}

\noindent
Proof. From Lemma~\ref{projinstrip}, $K$ belongs to the side~$5$ of~$P$. 
Let~$U$ and~$V$ be the reflections of~$P$ in~$\ell_2$ and~$\ell_3$ respectively.
From Lemma~\ref{visilow}, the lower strip of~$P$ 
is both contained in 
the lower strip of~$U$ and in that of~$V$. Accordingly,
$F$ belongs to side~$2$ and $G$ belongs to side~$3$. As $M$ is not in the same side 
of~$\ell_5$ as~$P$, $MF$, $MG$ cuts side~$5$ in~$R$, $S$ respectively: 
Note that, as $\ell_4$ is the common 
perpendicular to~$\ell_5$ and~$\ell_3$, and as $MK$ is perpendicular to the side~5
of~$P$, \hbox{$K\not=E$},  \hbox{$K\not=A$}, \hbox{$F\not=D$} and 
\hbox{$G\not=B$}. Also note that \hbox{$R\not=K$} and \hbox{$S\not=K$}. Otherwise, 
if~$R$ or~$S$ would coincide with~$K$, $KFBA$ or $DEKG$ respectively would be a 
rectangle, which is impossible. Now, by construction, $RFBA$ is a Lambert quadrangle, so 
that \hbox{$(RA,RF)$} must be acute. Clearly, \hbox{$(RM,RK)=(RA,RF)$}. As $MK$ is 
perpendicular to the side~5 of~$P$, \hbox{$(RM,RK)$} is an acute angle, so that we 
must have $[ARK]$: $R$ is inside $AK$. A similar argument with the Lambert quadrangle 
$EDGS$ shows us that $S$ is in $]KE[$. The case when~$M$ is on~$\ell_1$ or 
on~$\ell_4$ is obvious. \qed
\vskip 5pt
Let us go back to the cornucopia decomposition of~$\cal Q$. 

\begin{lem}\label{heads}
Consider the cornucopia decomposition of~$\cal Q$. Consider a region~$R$ of the
generation~$n$: let~$P_i$'s be the pentagons of the cornucopia of~$R$, and 
let~$R_i$'s be the regions of the generation $n$$+$$1$ inside~$R$, both sequences of 
objects being numbered as in the cornucopia of~$\cal Q$. Then, the head of the 
region~$R_0$ inside~$R$ is the image of the head of~$P_0$ by the shift along the 
side~$4$ of~$P_0$ with an amplitude of~$a$, and the head of the region~$R_i$ 
inside~$R$ with \hbox{$i\geq 1$} is the image of~$P_{i-1}$ under the shift along the 
side~$1$ of~$P_{i-1}$ with an amplitude of~$a$. Under these shifts, 
the correspondence between the sides/lines of~$P_i$ and those of the head 
of~$R_{i+1}$ as well as between the sides/lines of~$P_0$ and those of the head of~$R_0$ is given by Table~{\rm\ref{corressides}}.
\vskip -10pt
\newdimen\lalarge\lalarge=30pt
\newdimen\largetab\largetab=195pt
\def\cotes #1 #2 #3 #4 #5 #6 {%
\ligne{\rm%
\hbox to \lalarge {\hfill#1\hfill}
\hbox to \lalarge {\hfill#2\hfill}
\hbox to \lalarge {\hfill#3\hfill}
\hbox to \lalarge {\hfill#4\hfill}
\hbox to \lalarge {\hfill#5\hfill}
\hbox to \lalarge {\hfill#6\hfill}\hfill}
}
\setbox110=\vtop{\offinterlineskip\parindent=0pt\hsize=190pt
\cotes {$P_i$} 1 2 3 4 5
\cotes {$R_{i+1}$} 1 5 4 {} {} 
\cotes {$R_0$} {} 1 5 4 {}  
}
\vtop{\rm
\begin{tab}\label{corressides}\leurre
The numbers concern the lines when they are identical in~$R_{i+1}$ or~$R_i$ with 
those of~$P_i$.
\end{tab}
\vskip-5pt
\ligne{\hfill
\box110
\hfill}
}

\end{lem}

\noindent
Proof.
The line in Table~\ref{corressides} associated to~$R_0$ is a corollary of Lemma~\ref{visilow}.
For the regions~$R_i$ with \hbox{$i\geq 1$}, this is a corollary of Lemma~\ref{visi_shift}.
Remember 
that the shifts described in the statement of the lemma keep the orientation of the
numbering invariant and that due to the definition of the shifts, a side~$i$ is 
transformed into a side~$i$ under a shift along the support of the former side~$i$
for \hbox{$i\in\{1,4,5\}$}.
\qed
 
\begin{cor}\label{Ovisible}
Let~$R$ be a region in the cornucopia decomposition of~$\cal Q$. Let~$T$ be the head
of~$R$ and~$\ell$ be the line which supports the side~$5$ of~$T$. Then,
the half-plane defined by~$\ell$ which does not contain~$T$ contains~$O$.
\end{cor}

\noindent
Proof.
This is a corollary of Lemma~\ref{heads}. We know that the head of the region is
delimited by its side~$5$. From Lemma~\ref{projinstrip}, $O$ belongs to the lower strip 
of~$T_0$ and of~$T_1$, the heads of the region $R_0$ and~$R_1$ of generation~1. 
Lemma~\ref{visi_shift}
extends this property to all the other regions~$R_i$ of generation~1.

Assume that the property is true for the generation~$n$. Consider a region~$R^n$ of the 
generation~$n$. Let~$T$ be its head and let~$H$ 
be the half-plane defined by the support of the side~5 of~$T$ which does not 
contain~$O$. Then,  the heads of the regions $R_0$ and~$R_1$ of the 
generation $n$+1 are contained in~$H$, so that Lemma~\ref{heads} applied to~$T$ 
says that $O$~is also in the lower strip of the heads of the regions $R^{n+1}_0$ and~$R^{n+1}_1$ 
of the generation $n$+1: consequently,  the property also holds for these two regions. 
The shift along the side~5 of~$T$ of amplitude~$a$ which transforms the side~4 of~$T$
into its side~1 satisfies the hypothesis of Lemma~\ref{visi_shift}.
By induction, the lemma allows us to extend the property from
the region~$R^{n+1}_i$ with \hbox{$i\geq 1$} to the regions~$R^{n+1}_{i+1}$. 
Accordingly, the property is true for all regions of the generation~$n$+1. This completes 
the proof of the corollary. \qed

\begin{cor}\label{Oposit}
Consider a region~$R$ of the cornucopia decomposition of~$\cal Q$.
Then $O$~is in the lower strip of the head of~$R$. For another pentagon~$Q$ of the cornucopia of~$R$, $O$~is in the lower strip of the pentagon which is the image of~$Q$ under the shift along its side~$1$, going from the border of~$R$ to the side~$2$ 
of~$Q$.
\end{cor}

\noindent
Proof. This is also a consequence of the proof given for Corollary~\ref{Ovisible}.
\qed

\vskip 5pt
We arrive to the key property of this section.

\begin{lem}\label{cornudist}
In $\cal Q$, the distance from~$O$ to a region of the generation~$n$ is at least $n\cdot a$. 
\end{lem}

We need a preliminary result:

\begin{lem}\label{projsofO}
For each region~$R$ in the cornucopia decomposition of~$\cal Q$, the orthogonal 
projection of~$O$ on the border of~$R$ occurs on the side~$5$ of its head,
ends of the side excepted when $q$~is not a border of the region.
\end{lem}

\noindent
Proof. This is a corollary of Corollary~\ref{Oposit} and of Lemma~\ref{projpenta} and of 
the fact that the projection of~$ O$ on a region~$R$ is the same as its 
projection on the head~$H$ of~$R$: the projection is also the projection of~$O$ on the 
line~$\ell$ supporting the side~5 of~$H$. Accordingly, all points in the half-plane defined
by~$\ell$ which does not contain~$O$ are further from~$O$ than its projection 
on~$\ell$.\qed
\vskip 5pt
\noindent
Proof of Lemma~\ref{cornudist}.
 Note that the result is true for generation~1. The cornucopia of~$\cal Q$
has a complex border: it is $p$ which contains the side~5 of all pentagons contained
in the cornucopia. Another infinite part of the border consists of the sides~5 of the
heads of the regions of generation~1. As the pentagons~$P_i$ with
\hbox{$i\geq 1$} are outside the half-plane defined by the side~1 of~$P_0$ which does not contain~$O$, the distance of each~$P_i$ to~$O$ is at least~$a$. In particular, 
this is the case for $OK_n$ where $K_n$ is the orthogonal projection of~$O$ on the 
border of~$R_n$, \hbox{$n\geq 0$}. Now, as $R_i$ is contained in the half-plane defined
by the side~5 of its head containing its head, the distance from~$O$ to~$R_n$ is
at least~$OK_n$, so that it is at least~$a$.

Assume that the result is true for the generation~$n$. Consider a region~$R$ of the 
generation~$n$ and consider a region~$R^1$ of the generation~$n$+1 contained 
in~$R^1$. The head of~$R^1$ is obtained from a pentagon~$P_i$ of the cornucopia of~$R$. Now, from Lemma~\ref{projsofO}, the orthogonal projection~$K^1$ of~$O$ 
on~$R^1$ occurs on the head of~$R^1$. Let~$K$ be the orthogonal projection of~$O$ 
on$R$. Unless $R^1$ is the region~0 of~$R$, the head of~$R^1$ is obtained from the 
head of~$R$ by a shift along the side~1 of~$R$. From Lemma~\ref{projpenta}, we have
that $K^1$ is in the side~2 of the head of~$K$: informally, $OK^1$ is to the left 
of~$OK$. Let $OK^1$ cuts the side~5 of the head of~$R$ at~$L$. As the quadrangle
$ABK^1L$ is a Lambert quadrangle, remember that $AB$ is the side~1 of the head 
of~$R$, the angle~$(LA,LK^1)$ is acute, so that \hbox{$LK^1>a$}. On another hand,
\hbox{$OL>OK$} as~$L$ is on the side~5 of the head of~$R$ and as 
\hbox{$L\not=K$}. Accordingly, \hbox{$OK^1=OL+LK^1>n\cdot a+a$}.
If $R^1$ is the region~0, then $K^1$ is on the side~3 of the head of~$R$. Now, we consider the quadrangle $LK^1DE$ which is also a Lambert quadrangle, so that the same estimates can be performed, leading us to the same conclusion. And so, the property is 
true for the regions of the generation~$n$+1.
\qed

Lemma~\ref{cornudist} has a very important corollary which we establish now, although it is not tightly connected to the topic of this paper.

The left-hand side picture of Figure~\ref{tree_bij} illustrates the bijection between the 
restriction of the pentagrid to~$\cal Q$ with a tree we called the Fibonacci tree, 
see~\cite{mmJUCSii,mmbook1}. The name of the tree comes from the fact that the 
number of nodes of the tree which are at the same distance~$d$ from its root in term of 
crossed tiles is $f_{2d+1}$ where \hbox{$\{f_n\}_{n\in\mathbb{N}}$} is the Fibonacci 
sequence with \hbox{$f_0=f_1=1$}. In~\cite{mmbook1}, we remember the proof of the 
property already mentioned in~\cite{mmJUCSii,mmkmTCS} that the restriction of the 
pentagrid to~$\cal Q$ is in bijection with a tree which we called the Fibonacci tree:
The tree can be constructed by the infinite iteration of two rules we can formulate as 
$W\rightarrow BWW$ and $B\rightarrow BW$, $B$ denoting the nodes which have two 
sons and $W$ denoting those which have three of them, the root of the tree being 
a $W$-node.  We can state the following result:

\begin{thm}\label{thmbij} {\rm(see~\cite{mmkmTCS,mmJUCSii,mmbook1})}
The Fibonacci tree is in bijection with the restriction of the pentagrid to $\cal Q$.
\end{thm}

\noindent
Proof.
The proof of the injection is easy: it is enough to note that the sons of a node~$\nu$ are obtained by the reflection of the tile~$T$ associated to~$\nu$ in two or three
different sides of~$T$. 

For the surjection, we have to prove that any point of 
$\cal Q$ belongs to a tile of the pentagrid restricted to~$\cal Q$. Using the cornucopia 
decomposition, it is rather easy. Let $M$~be a point of~$\cal Q$. If~$M$ belongs to the 
cornucopia of~$\cal Q$, it belongs to some~$P_i$ and we are done. If this is not the 
case, it belongs to some~$R^1$ of generation~1. In~$R^1$ we repeat the same argument: either $M$
belongs to the cornucopia of~$R^1$ and we find a pentagon of the tiling containing~$M$, or we find that $M$ belongs to some region~$R^2$ of generation~2.
As from Lemma~\ref{cornudist}, the distance from~$O$ to a region of the generation~$n$ is at least~$n\cdot a$, we can find an~$m$ such that
$m\cdot a>OM$, so that necessarily, $M$ belongs to the cornucopia of a region~$R^k$
of the generation~$k$ with $k<m$. Eventually, $M$ belongs to some pentagon of the tiling. 

It is not difficult to see that the pentagons of the cornucopia decomposition are those of the Fibonacci tree: the cornucopia of~$\cal Q$ corresponds to the leftmost branch of the Fibonacci tree. Its regions $R_0$ and~$R_1$ have the $W$-sons of the root for
their respective heads. Now, in each region, the cornucopia is the leftmost branch of the sub-tree rooted at the node corresponding to the head.  Note that the heads of the regions are the white nodes of the tree, the root being the head of~$\cal Q$. This is illustrated by 
the right-hand side picture of Figure~\ref{tree_bij}.\qed

\vtop{
\vspace{10pt}
\ligne{\hfill\includegraphics[scale=0.7]{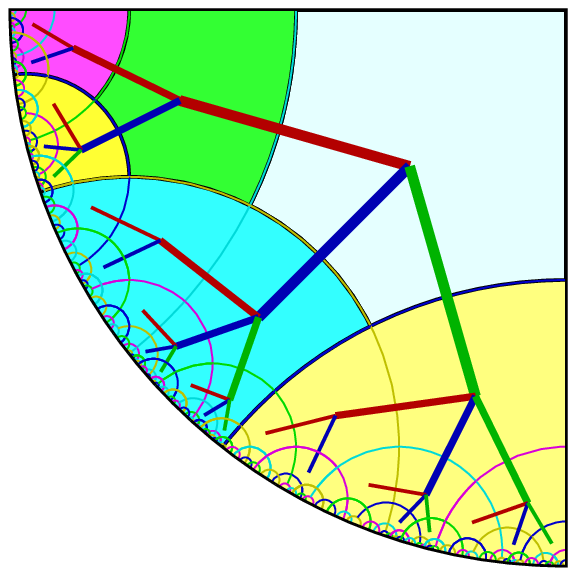}
\hfill
\raise-5pt\hbox{\includegraphics[scale=0.75]{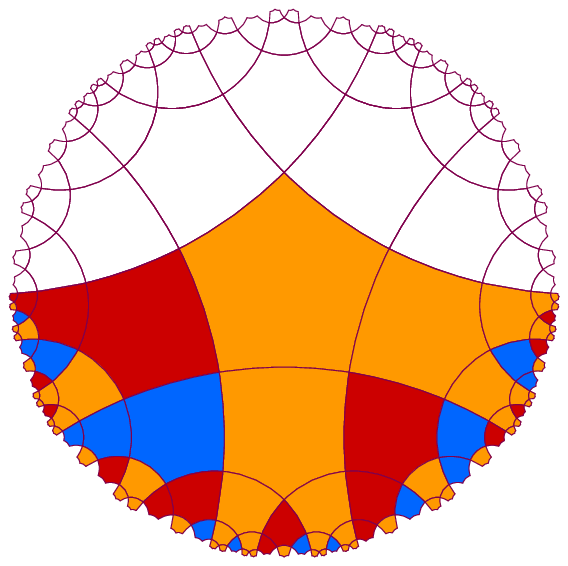}}
\hfill}
\vspace{-5pt}
\begin{fig}
\label{tree_bij}
\leurre
To left:
the bijection between the tree and the quarter. A red arrow leads to a \textbf{black node},
the others lead to a \textbf{white} one. The root of the tree is considered as a white node.
\vskip 0pt
\noindent
To right: correspondence between the cornucopias of the decomposition of~$\cal Q$ and the black nodes of the Fibonacci tree. The black nodes are the tiles in blue and in red. The other coloured tiles are white nodes.
\end{fig}
}

\section{Convergence at infinity}
\label{convinf}

   In the following sections, we shall have to deal with sequences of points which are
converging to infinity. Convergence in the hyperbolic plane is easy and we can rely on
Poincar\'e's disc model as far as topology only is concerned. To study points at infinity,
we have to resist to the use of Poincar\'e's disc model: it fairly represents what Hilbert
called ends in the hyperbolic plane, but there is always the danger that the Euclidean
intuition plays some bad trick on us. In order to define convergence to infinity,
we have to justify that the notion of convergence to a point of the border in 
Poincar\'e's disc model turns out to be valid.

Consider the same fixed point~$O$ of the hyperbolic plane and the same 
quarter~$\cal Q$ whose vertex is~$O$ which were defined in Section~\ref{cornucopia}
and consider a sequence~$\sigma$ of points 
\hbox{\{$x_n\}_{n\in\mathbb{N}}$} of the hyperbolic plane such that \hbox{$x_n\in\cal Q$}
for all~$n$ and that
$Ox_n$ tends to infinity as $n$~tends to infinity. Say that a \textbf{neighbourhood of infinity}
for~$\sigma$ is a half-plane~$H$ defined by a line~$\ell$ such that
\hbox{$\mathbb{H}^2\backslash H$} contains finitely many points of~$\sigma$ only.

Consider a point at infinity~$\alpha$ and a line~$\ell$ which does not pass 
through~$\alpha$. The line defines two half-planes:
$H_1$ and~$H_2$. In one of them, say~$H_2$, any line contained in the half-plane does not pass 
through~$\alpha$. In the other, there are such lines~: for any point~$M$ of~$\ell$ there is a 
unique line which passes through~$M$ and through~$\alpha$. In our sequel we say that $H_1$ is the
half-plane defined by~$\ell$ which \textbf{touches}~$\alpha$ and that~$H_2$ is the one which 
does \textbf{not touch}~$\alpha$.

\begin{lem}\label{faraway}
Let $\ell$ be a line of the hyperbolic plane and let~$H$ be the half-plane delimited by~$\ell$
which does not contain~$O$. Let~$K$ be the orthogonal projection of~$O$ on~$\ell$.
Let~$\delta_1$ and $\delta_2$ be the ray issued from~$O$ which are parallel to~$\ell$.
Then \hbox{$(\delta_1,\delta_2)$} tends to zero as $OK$ tends to infinity and conversely.
\end{lem}

\noindent
Proof. By construction, as \hbox{$OK\perp\ell$}, \hbox{$(\delta_1,OK)$} is the angle of 
parallelism of~$OK$ for~$\ell$. The conclusion of the lemma is a well known property
already established by Lobachevsky.\qed

\begin{lem}\label{infneigh}
Let $\alpha$ be a point at infinity. Let~$\delta^1_n$ and 
$\delta^2_n$ be two rays issued from~$O$ such that $O\alpha$ is the bisector of the angle
\hbox{$(\delta^1_n,\delta^2_n)$} and \hbox{$(\delta^1_n,\delta^2_n)<\displaystyle{{\pi}\over n}$}. 
Then, there is a unique line $\ell_n$ 
of the hyperbolic plane such that $\ell_n$ is parallel to both~$\delta^1_n$ and~$\delta^2_n$.
Let $H_n$ be the half-plane defined by~$\ell_n$ which does not contain~$O$. Then
the $H_n$'s constitute a basis of neighbourhoods for~$\alpha$.
\end{lem}

\noindent
Proof. The existence of~$\ell_n$ is a well known property: it comes from the fact that $\ell_n$
is the unique line of the hyperbolic plane which is parallel to~$\delta_1$ and which is perpendicular to~$O\alpha$.
In order to prove that the $H_n$'s constitute a basis of neighbourhoods
for~$\alpha$, we first note that a neighbourhood of~$\alpha$ is a subset of
$\mathbb{H}^2$ which contains a half-plane~$H$ which touches~$\alpha$. Of course, we may
assume that~$H$ does not contain~$O$. Now, let~$\ell$ be the border of~$H$.
Consider the ray~$O\alpha$: it cuts~$\ell$ at~$K$, otherwise, $H$ cannot touch~$\alpha$. 
If it is perpendicular to~$\ell$, there is a point~$L$ 
on~$O\alpha$ with $[OKL]$ such that the parallel~$\ell_n$ issued from~$L$ to~$\delta^1_n$ is
perpendicular to~$O\alpha$. Then, as $O\alpha$ is the bisector of
\hbox{$(\delta^1_n,\delta^2_n)$}, $\ell_n$ is also parallel to~$\delta^2_n$.
 
If $O\alpha$ is not perpendicular to~$\ell$, then there is a point~$L$ on~$O\alpha$
with $[OKL]$ such that the perpendicular~$\mu$ to~$O\alpha$ passing through~$L$ is non-secant 
with~$\ell$. We repeat with~$\mu$ the just above argument.

We remain to prove that if $\beta$ is another point at infinity, so that 
\hbox{$\beta\not=\alpha$}, there is a $H_n$ so that $H_n$ does not touch~$\beta$: we may even
construct $H_n$ so that its border~$\ell_n$ does not pass through~$\beta$.
Indeed, we take~$n$ so that \hbox{$\displaystyle{1\over n}<(O\alpha,O\beta)$} and we repeat
the above construction. It is then plain that $H_n$ is contained in the half-plane delimited
by~$O\delta^1_n$ which contains~$\alpha$. By the construction, this latter half-plane
does not touch~$\beta$ as $O\delta^1_n$ does not pass through~$\beta$ and as we may assume
that $\alpha$ and~$\beta$ are not on the same side of~$O\delta^1_n$.\qed
\vskip 5pt
Say that a line of the hyperbolic plane is a \textbf{line of the pentagrid} if it supports
at least an edge of a pentagon of the tessellation. We wish to prove that in Lemma~\ref{infneigh} we can replace the lines~$\ell_n$ by lines of the pentagrid.
To this aim we prove the following result.

\begin{lem}\label{linepenta}
Let $\alpha$ be a point at infinity of the hyperbolic plane and let~$\ell$ be a line
which does not pass through~$\alpha$. Then there is a line of the pentagrid~$\lambda$ such
that $\lambda$ is completely contained in the half-plane defined by~$\ell$ which
touches~$\alpha$.
\end{lem}

\noindent
Proof. 
Let~$b$ be the diameter of the regular rectangular pentagon.
It is plain that \hbox{$a<b<\displaystyle{5\over 2}a$}: take any picture in 
Figure~\ref{bij} to check the latter inequality as $b$ is the distance from a vertex
of the pentagon to the midpoint of the opposite side. This means that for any point~$P$ of the
hyperbolic plane, within a disc of radius~$b$ centered at~$P$ we can find a vertex of the 
pentagrid. Consider~$\ell$, a line of the hyperbolic plane which does not pass through~$\alpha$.
Denote by~$H_1$ the half-plane defined by~$\ell$ which touches~$\alpha$ and by~$H_2$
the other half-plane: that which does not touch~$\alpha$. Take~$A$ a point on~$\ell$ and let
$S$~be a vertex of the pentagrid such that \hbox{$AS\leq b$}, which is in~$H_1$ and which is 
the closest to~$S$. Let $r_1$ and~$r_2$ be the rays issued from~$S$ which are supported by the 
lines of the pentagrid which meet at~$S$ and which delimit a quarter whose head~$P$ cuts~$\ell$. 
If both~$r_1$ and~$r_2$ do not meet~$\ell$, we are done. If we require~$r_1$ and~$r_2$
to be non-secant with~$\ell$, we take $R$ on the continuation of~$r_1$ in~$H_1$, at the distance~$a$
and then we take~$T$ on the next side of the pentagon~$Q$ which contains~$R$ and~$S$ and 
which has a common side of~$P$. Then the rays issued from~$T$ and supporting the edges of~$Q$
abutting~$T$ are non-secant with~$r_1$ and~$r_2$ as having a common perpendicular with these
rays. At least one of the half-planes delimited by~$r_1$ and~$r_2$ touches~$\alpha$. We take the line corresponding to this half-plane.

    If $r_1$ and~$r_2$ are not in this case, at least one of them, say~$r_1$ cuts~$\ell$.
Continue the ray~$r_1$ by the other ray~$y_1$ on the same line
until we meet a vertex~$R$ of the pentagrid for which the other ray~$r_3$ abutting~$R$ and which is on the same side of~$r_1$ as~$P$, is non-secant with~$\ell$. 
Indeed, let~$K$ be the orthogonal projection of~$R$
on~$\ell$. As $R$ tends to infinity on~$y_1$, $RK$ also tends to infinity and the angle
of~$y_1$ with~$RK$ tends to zero so that we can find such an~$R$ that the angle of parallelism
for~$RK$ with~$\ell$ is less than $\displaystyle{\pi\over 4}$. Then the angle~$\vartheta$ 
of~$y_1$ with~$RK$ satisfies \hbox{$\vartheta<\displaystyle{\pi\over 4}$} as $r_1$ 
cuts~$\ell$. Accordingly $r_3$ makes an angle which is bigger than $\displaystyle{\pi\over 4}$
so that $r_3$ and its continuation in a line is non-secant with~$\ell$ and it clearly
lies in~$H_1$. Let~$y_3$ be the continuation of~$r_3$ after~$R$. Then, we can find on~$y_3$ a 
vertex~$T$ of the pentagrid so that the perpendicular~$y_4$ to~$y_3$ passing through~$T$ is 
non-secant with~$\ell$. Then at least one of the half-planes delimited by~$y_3$ and~$y_4$ and 
which does not contain~$\ell$ touches~$\alpha$. We take the line defined by this half-plane.
\qed

\begin{cor}\label{neighgrid}
The lines of the pentagrid define neighbourhoods for the points at infinity.
\end{cor}

\noindent
Proof. It is a direct consequence of Lemmas~\ref{infneigh} and~\ref{linepenta}.
\qed

\section{Preliminary properties}\label{prelim}

   Figure~\ref{decomp} indicates two ways to decompose a quarter into other quarters.

Consider two quarters~$F_1$ and~$F_2$ whose vertices are~$S_1$ and~$S_2$ respectively and
whose heads are $H_1$ and~$H_2$ respectively. We say that $F_1$~is \textbf{embedded},
\textbf{strictly embedded} in~$F_2$, denoted by \hbox{$F_1\sqsubseteq F_2$}, 
\hbox{$F_1\sqsubset F_2$} respectively, 
if \hbox{$F_1\subseteq F_2$}, \hbox{$F_1\subset F_2^\circ$} respectively,
where $F_2^\circ$ is the interior of~$F_2$.
From the definition, strictly embedded quarters are embedded but embedded quarters may be
not strictly embedded. Denote by $\partial F$ the border of the quarter~$F$.
In the left-hand side of Figure~\ref{decomp}, we can see that the orange quarter is embedded
in the quarter~$Q$ whose head is the red tile. We also can see on the same picture
that the blue quarter is strictly embedded in~$Q$. On the right-hand side of Figure~\ref{decomp},
the blue quarter and the quarter which extends the light orange zone are both embedded in~$Q$, 
but not strictly.
In the situation when the head of~$F_1$ shares an edge with the head of~$F_2$,
there are three possible cases. In two of them, $F_1$ is embedded in~$F_2$ but not
strictly, while in the third case, $F_1$ is strictly embedded in~$F_2$. We shall
denote these cases by \hbox{$F_1\sqsubseteq_0 F_2$} when the embedding is not strict
and \hbox{$F_1\sqsubset_0 F_2$} when the embedding is strict. 
The index~0 reminds us that the heads share an edge. In both cases we speak of
a \textbf{one step} embedding. Note that when \hbox{$F_1\sqsubset_0 F_2$}, it is not possible
to find a quarter~$G$ such that \hbox{$F_1\sqsubseteq_0 G$} and
\hbox{$G\sqsubseteq_0 F_2$}.  Now, we can prove the property indicated in
Lemma~\ref{dist_embedded}.

\vtop{
\vspace{10pt}
\ligne{\hfill\includegraphics[scale=0.7]{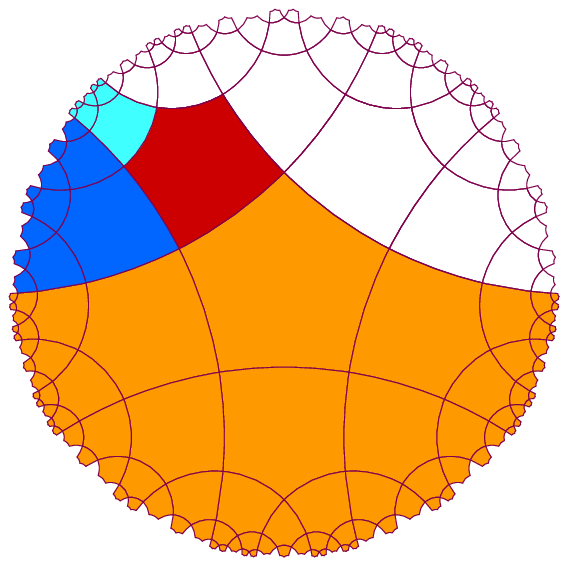}
\hfill\includegraphics[scale=0.7]{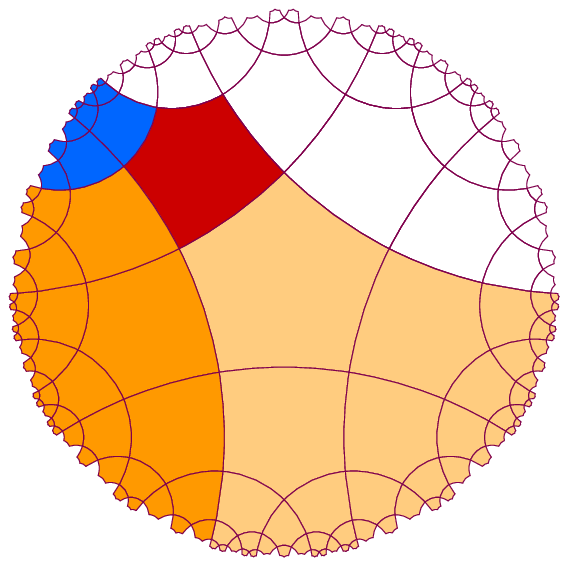}
\hfill}
\vspace{-5pt}
\begin{fig}
\label{decomp}
\leurre
To left: the left-hand side decomposition.
\vskip 0pt
\parindent 0pt
To right: the central decomposition.
\end{fig}
}

\begin{lem}\label{dist_embedded}
Let $F_1$ and $F_2$ be two embedded quarters whose vertices are~$S_1$ and~$S_2$
respectively. There is a finite sequence
\hbox{$G_1,\ldots,G_k$} of quarters such that \hbox{$F_1=G_1$}, 
\hbox{$F_2=G_k$} and \hbox{$G_i\sqsubset_0 G_{i+1}$} 
or \hbox{$G_i\sqsubseteq_0 G_{i+1}$} for \hbox{$i\geq 1$ and $i<k$}. Moreover,
the distance from~$S_1$ to~$S_2$ is $k$$-$1{} in number of tiles. 
\end{lem}

\ifnum 1=0 {
\vtop{
\ligne{\hfill\includegraphics[scale=0.7]{embedded.ps}
\hfill\includegraphics[scale=0.7]{str_embedded.ps}
\hfill}
\vspace{-5pt}
\begin{fig}
\label{embedded}
\leurre
To right: two strictly embedded quarters. To left: two embedded quarters which are not
strictly embedded. 
\vskip 0pt
Note that in the right hand-side picture, the border of the bigger quarter is at distance~$a$
from the smaller one.
\end{fig}
}
} \fi

\noindent
Proof. Identifying the head of~$F_2$ as the tile in bijection with the root of the
Fibonacci tree, see the left-hand side picture of Figure~\ref{tree_bij}, it is easy to 
find a finite sequence
of tiles~$T_i$, with \hbox{$i\in[1..k]$}, with $T_1$ being the head of~$F_2$
and~$T_k$ that of~$F_1$. Each tile is in correspondence with the nodes of the tree
which are on the branch which leads from the root to the node in bijection with the head
of~$F_1$. By construction, $T_1$ is the head of $F_2$ and \hbox{$F_2=G_k$} by construction. 
For each~$T_i$ with $i>1$, we look at the place of~$T_i$ with respect to~$T_{i-1}$ which is
the head of~$G_{k-i+1}$. There are 
three possible cases only as indicated by Figure~\ref{decomp}. If the edge shared 
with~$T_{i-1}$ has a vertex on the border of~$G_{k-i+2}$, then
we define~$G_{k-i+1}$ as indicated by~Figure~\ref{decomp}: there is a single possibility
which yields \hbox{$G_{k-i+1}\sqsubseteq_0 G_{k-i+2}$}.
If the edge shared with~$T_{i-1}$ has no vertex on the border of~$G_{k-i+2}$, there
is again a single possibility given by the left-hand side decomposition and we
have \hbox{$G_{k-i+1}\sqsubset_0 G_{k-i+2}$}.
The distance in number of tiles from~$S_1$ to~$S_2$ is the number of tiles on the
branch, the last tile being excepted, so it is~$k$$-$1.\qed 

\begin{cor}\label{dist_vertex}
Let $F_1$ and $F_2$ be two quarters whose vertices are~$S_1$ and~$S_2$ respectively.
Then \hbox{dist$(S_1,S_2)>a$}. 
\end{cor}

\noindent
Proof.
Clearly, the distance is bigger if the embedding is not in one step. For a one step
embedding,
Figure~\ref{decomp} clearly proves Corollary~\ref{dist_vertex}.\qed

\section{Sequences of quarters}
\label{sequences}

From what we have seen in Section~\ref{prelim},
when we are dealing with a sequence \hbox{$\{F_n\}_{n\in\mathbb{N}}$} of quarters
such that \hbox{$F_n\sqsubseteq F_{n+1}$}, we may assume that each embedding of consecutive
terms of the sequence is a one step embedding. Say that such a sequence is \textbf{stepwise}.

Now, consider a stepwise sequence $\{F_n\}_{n\in\mathbb{N}}$ of embedded quarters. 
Let $S_n$ be the vertex of~$F_n$.
The sequence \hbox{$\{S_n\}_{n\in\mathbb{N}}$} cannot converge in the hyperbolic plane
as the distance between two consecutive terms is at least~$a$. Note that the topologies
induced in Poincar\'e's disc by the Euclidean metric and by the hyperbolic one coincide
despite the fact that the metrics are very different. This is a well known feature, coming from 
the property that hyperbolic circles are Euclidean circles contained in the open disc. Now, the 
closure of the disc is compact, so that the sequence \hbox{$\{S_n\}_{n\in\mathbb{N}}$}
has at least one limit point~$\alpha$ which is a point of the border of the Poincar\'e's
disc, which corresponds to an end of the hyperbolic plane. 

Consider three consecutive terms of the sequence: $F_n$, $F_{n+1}$ and $F_{n+2}$.
Consider the one-step relations between consecutive terms. If we have both
\hbox{$F_n\sqsubseteq_0 F_{n+1}$} and \hbox{$F_{n+1}\sqsubseteq_0 F_{n+2}$}, 
we have two cases: we have either 
\hbox{$F_n\sqsubset F_{n+2}$} or \hbox{$F_n\not\sqsubset F_{n+2}$}, but in that latter
case, we also have  \hbox{$F_n\sqsubseteq F_{n+2}$}, see the first two pictures
of Figure~\ref{onesteplarge}.
The figure shows us that starting from $F_{n+1}$, there are two possibilities
to construct $F_{n+2}$ and only them: those which are illustrated by the pictures of the
figure. Indeed, we have only two possibilities for choosing the new head. Once the new head 
is chosen, we have \textit{a priori} two possibilities for choosing the vertex in order
to obtain a quarter which contains $F_{n+1}$. But one of them defines $F_{n+2}$ as strictly
embedding $F_{n+1}$. So that a single vertex remains to obtain $F_{n+2}$ as embedding
$F_{n+1}$ but not strictly, see the first two pictures of Figure~\ref{onesteplarge}. And so, 
we remain with the two cases which are illustrated by the first two pictures of
Figure~\ref{onesteplarge}.

\vtop{
\vspace{10pt}
\ligne{\hfill
\includegraphics[scale=0.65]{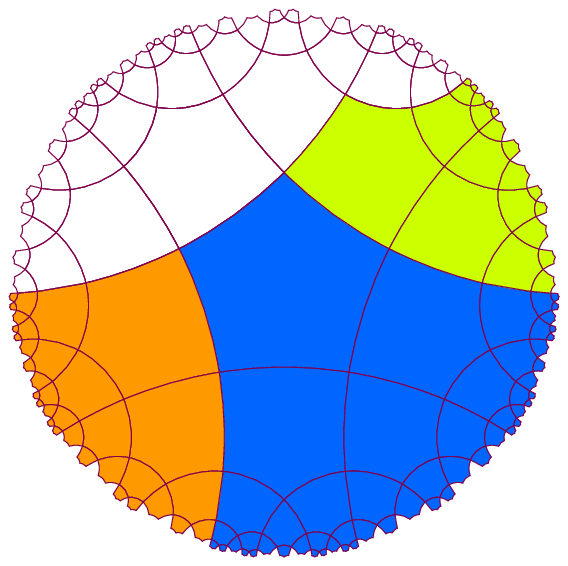}
\hfill
\includegraphics[scale=0.65]{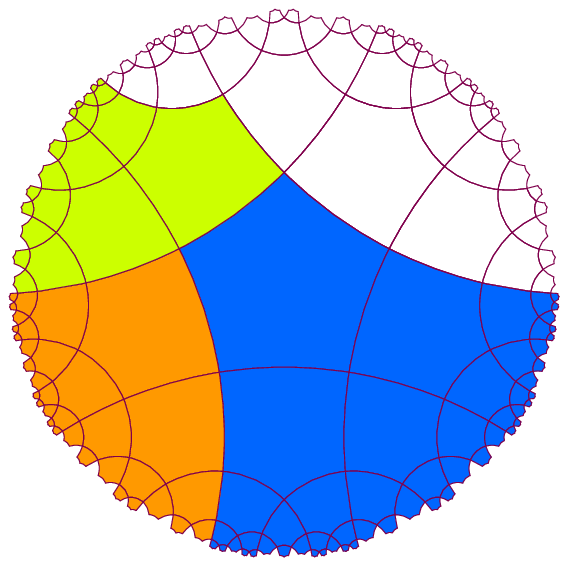}
\hfill
\includegraphics[scale=0.65]{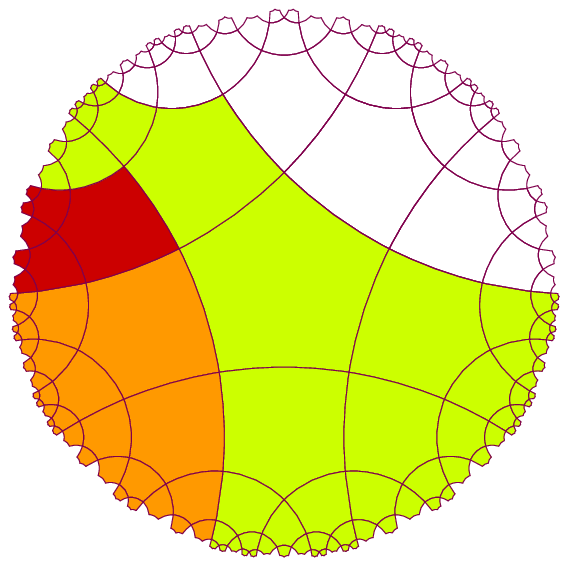}
\hfill}
\vspace{-5pt}
\begin{fig}
\label{onesteplarge}
The cases when \hbox{$F_n\sqsubseteq_0 F_{n+1}$} and \hbox{$F_{n+1}\sqsubseteq_0 F_{n+2}$}.
\vskip 0pt
\noindent
To left: we have that \hbox{$F_n\sqsubseteq F_{n+2}$} and \hbox{$F_n\not\sqsubset F_{n+2}$}.
\vskip 0pt
\noindent
Centre: we have that \hbox{$F_n\sqsubset F_{n+2}$}. 
\vskip 0pt
\noindent
To right: we have \hbox{$F_n\sqsubseteq_0 F'_{n+1}$} and \hbox{$F'_{n+1}\sqsubset_0 F_{n+2}$}
\end{fig}
}

\subsection{Non-alternating sequences}
\label{non_altern}

Now, the rightmost picture of Figure~\ref{onesteplarge} show us the following property:

\begin{lem}\label{no_altern}
Consider three quarters, $F_1$, $F_2$ and~$F_3$ such that 
\hbox{$F_1\sqsubseteq_0 F_2$} and
\hbox{$F_2\sqsubset_0 F_3$}. 
Then, there is a quarter~$F_4$ such that:
\hbox{$F_1\sqsubseteq_0 F_4$} and
\hbox{$F_4\sqsubseteq_0 F_3$}. Conversely, if we have
\hbox{$F_1\sqsubseteq_0 F_2$}, 
\hbox{$F_2\sqsubseteq_0 F_3$} and
\hbox{$F_1\sqsubset F_3$}, we may find~$F_4$ such that
\hbox{$F_1\sqsubseteq_0 F_4$} and
\hbox{$F_4\sqsubset_0 F_3$}.

\end{lem}

\begin{cor}\label{large_and_strict}
Let \hbox{$\{F_n\}_{n\in\mathbb{N}}$} be a stepwise sequence of consecutively embedded quarters.
We may assume that if
\hbox{$F_n\sqsubseteq_0 F_{n+1}$}, 
\hbox{$F_{n+1}\sqsubseteq_0 F_{n+2}$} and 
\hbox{$F_n\sqsubset F_{n+2}$}, then we have
\hbox{$F_{n+1}\sqsubset_0 F_{n+2}$}.
\end{cor}

Consider a stepwise  sequence of embedded quarters \hbox{$\{F_n\}_{n\in\mathbb{N}}$}. 
Say that $F_{n+1}$ presents an {\bf alternation} if and only if
\hbox{$F_n\sqsubseteq_0 F_{n+1}$}, \hbox{$F_{n+1}\sqsubseteq_0 F_{n+2}$} and
\hbox{$F_n\sqsubset F_{n+2}$}. From Corollary~\ref{large_and_strict}, we may assume
that a stepwise sequence \hbox{$\{F_n\}_{n\in\mathbb{N}}$} does not contain any alternation.
This necessarily means that if \hbox{$F_n\sqsubset F_{n+2}$}, then 
\hbox{$F_{n+1}\sqsubset_0 F_{n+2}$}. We say that a stepwise sequence of
embedded quarters \hbox{$\{F_n\}_{n\in\mathbb{N}}$} with no-alternation is
\textbf{ultimately direct} if there is an integer~$N$ such that for
all positive~$k$ we have \hbox{$F_N\not\sqsubset F_{N+k}$}. 
If in an ultimately direct sequence we may have \hbox{$N=0$}, we say that the sequence is
\textbf{direct}. We can state:

\begin{lem}\label{non_alternate}
Let \hbox{$\{F_n\}_{n\in\mathbb{N}}$} be a stepwise sequence of embedded quarters with
no alternation and assume the sequence to be ultimately direct.
Let~$S_n$ be the vertex of~$F_n$. Then $S_n$ tends to the point at infinity~$\alpha$
which is on a line~$\ell$ which supports one border of all $F_n$'s starting from a certain rank.
Moreover, all $F_n$'s are contained in the same half-plane defined by~$\ell$.
\end{lem}

\noindent
Proof. From the assumption, we have an integer~$N$ such that
\hbox{$F_n\sqsubseteq_0 F_{n+1}$} for all~$n\geq N$ and such that 
\hbox{$F_N\not\sqsubset F_{N+k}$} for any positive~$k$. And so, there is a line~$\ell$
issued from~$S_N$ such that $\ell$ contains a part of the border of~$F_N$ and such that
for all positive~$k$, there is a ray issued from~$S_{N+k}$ which is in the border of~$F_{N+k}$
and which is contained in~$\ell$. Clearly, $S_{N+k}$ converges to a point at infinity which
is on~$\ell$ as $k$ tends to infinity. Also clearly, as there is no alternation,
all $F_{N+k}$'s are on the same side of~$\ell$. Due to the consecutive embedding of
all terms of the sequence, all $F_n$'s are also in the same side.\qed

\subsection{Limit of vertices}
\label{limvert}

Now, what can be said for stepwise sequences of embedded quarters with no alternation
which are not ultimately direct?

\begin{lem}\label{close}
Consider a stepwise sequence $\{F_n\}_{n\in\mathbb{N}}$ of embedded quarters and assume
it to be with no alternation and assume that the sequence is not ultimately
direct. 
Let~$S_n$ be the vertex of~$F_n$.
Assume that the sequence \hbox{$\{S_n\}_{n\in\mathbb{N}}$} converges to an end~$\alpha$.
Let $\ell$ be a line which does not pass through~$\alpha$. Then, there is an~$N$ such
that for all~$n$, \hbox{$n\geq N$}, $F_n$ contains the half-plane delimited by~$\ell$
which does not touch~$\alpha$.
\end{lem}

Assuming Lemma~\ref{close}, we can prove:

\begin{thm}\label{to_end}
Consider a sequence $\{F_n\}_{n\in\mathbb{N}}$. Let $S_n$ be the vertex of~$F_n$. Then
there is an end~$\alpha$ such that $S_n$ converges to~$\alpha$ when $n$~tends to infinity.
\end{thm}

\noindent
Proof of Theorem~\ref{to_end}. From what we have already noticed, the sequence
\hbox{$\{S_n\}_{n\in\mathbb{N}}$} has at least one limit point, and any limit point is
an end. Assume that the sequence has at least two distinct limit points~$\alpha_1$
and~$\alpha_2$. Then we can find lines~$\ell_1$ and~$\ell_2$ such that if $\pi_1$ and $\pi_2$
respectively are the half-planes defined by~$\ell_1$ and~$\ell_2$ and which touches
$\alpha_1$ and~$\alpha_2$ respectively, then \hbox{$\pi_1\cap\pi_2=\emptyset$}. 
Indeed, consider two lines $m_1$ and $m_2$ which pass by $\alpha_1$ and $\alpha_2$
respectively. As \hbox{$\alpha_1\not=\alpha_2$}, the lines are distinct. We may assume that they
meet at some point~$A$ of the hyperbolic plane. If not, the lines are non secant. Then
replace~$m_1$ and~$m_2$ by the lines which are parallel to $m_1$ and~$m_2$
and which are issued from the mid-point of the segment of the
common perpendicular to~$m_1$ and~$m_2$ which joins~$m_1$ to~$m_2$. From~$A$, consider the
bisector of the angle \hbox{$(A\alpha_1,A\alpha_2)$}. It defines a point at infinity~$\beta$.
Then take the bisector of \hbox{$(A\alpha_1,A\beta)$} and of \hbox{$(A\beta,A\alpha_2)$}.
These new bisectors define two new points at infinity~$\beta_1$ and~$\beta_2$. Now,
define~$\ell_1$, $\ell_2$ as the perpendicular to~$m_1$, $m_2$ respectively, issued from~$\beta_1$,
$\beta_2$ respectively.

Consider two sub-sequences of the $F_n$'s, \hbox{$\{G_k\}_{k\in\mathbb{N}}$}
and \hbox{$\{H_k\}_{k\in\mathbb{N}}$} such the vertices of the $G_k$ converge to~$\alpha_1$
and those of the $H_k$ converge to~$\alpha_2$. We have
\hbox{$G_k=F_{n_k}$} and \hbox{$H_k=F_{m_k}$} where \hbox{$\{n_k\}_{k\in\mathbb{N}}$}
and \hbox{$\{m_k\}_{k\in\mathbb{N}}$} are distinct sub-sequences of~$\mathbb{N}$.

Assume that \hbox{$\{G_k\}_{k\in\mathbb{N}}$} and \hbox{$\{H_k\}_{k\in\mathbb{N}}$} are
both ultimately direct. There is an integer~$N$ such that
for all positive~$k$ we have both \hbox{$G_N\sqsubseteq G_{N+k}$} and
\hbox{$H_N\sqsubseteq H_{N+k}$} together with both \hbox{$G_N\not\sqsubset G_{N+k}$}
and \hbox{$H_N\not\sqsubset H_{N+k}$}. There is a line~$m_g$ and a line~$m_h$
such that $m_g$ passes through $\alpha_1$ and $m_h$ passes through $\alpha_2$ and,
from Lemma~\ref{non_alternate}, all~$G_k$'s are on the same side of~$m_g$ and
all~$H_k$'s are on the same side of~$m_h$. These sides define half-planes ${\cal H}_g$,
${\cal H}_h$ delimited by $m_g$, $m_h$ respectively. Assume that
\hbox{${\cal H}_h$} contains \hbox{${\cal H}_g$} when we get close to~$\alpha_2$.
Consider some~$m$ with \hbox{$m>N$}. We can find \hbox{$n>m$} such that
we have for instance that $G_n$ contains $H_m$: but then, $G_n$ contains points of the hyperbolic
plane which are not in \hbox{${\cal H}_g$}, a contradiction. 
So that we now assume that ${\cal H}_h$ and ${\cal H}_g$ do not meet when we get close 
to~$\alpha_2$. But the same inclusion as above immediately shows that $G_n$ contains 
points which are not in ${\cal H}_g$.  Accordingly, in that case,
\hbox{$\alpha_1=\alpha_2$}.

Consider now that one \hbox{$\{G_k\}_{k\in\mathbb{N}}$} is ultimately direct, and that
\hbox{$\{H_k\}_{k\in\mathbb{N}}$} is not.
From Lemma~\ref{close} there is~$N$ such that when \hbox{$h\geq N$}, $H_h$ contains 
\hbox{$\mathbb{H}^2\backslash\pi_2$}. Now, assume that 
\hbox{${\cal H}_g\cap\pi_2=\emptyset$}. Take a~$k>N$ such that \hbox{$n_k>m_h$}.
Then, $H_h$ contains \hbox{$\mathbb{H}^2\backslash\pi_2$} as well as a few points
of~$\pi_2$. As \hbox{$n_k>m_h$}, $G_k$ contains also points in~$\pi^2$,
a contradiction as \hbox{$G_k\subset {\cal H}_g$}. Now,
assume that \hbox{$\pi_2\subset {\cal H}_g$}. Again, take $k>N$ such that \hbox{$n_k>m_h$}. 
As \hbox{$G_k$} contains~$H_h$, it also contains points which are on the complement
of ${\cal H}_g$, again a contradiction. And so, in that case too, 
\hbox{$\alpha_1=\alpha_2$}.

Now, we remain with the case when both sequences \hbox{$\{G_k\}_{k\in\mathbb{N}}$}
and \hbox{$\{H_k\}_{k\in\mathbb{N}}$} are not ultimately direct.
From
Lemma~\ref{close} there is~$N$ such that when \hbox{$n\geq N$}, $H_n$ contains both
\hbox{$\mathbb{H}^2\backslash\pi_1$} and \hbox{$\mathbb{H}^2\backslash\pi_2$}. But
\hbox{$(\mathbb{H}^2\backslash\pi_1)\cup(\mathbb{H}^2\backslash\pi_1)=\mathbb{H}^2$},
so that $F_m$ contains~$\mathbb{H}^2$ for a certain~$m$, which is impossible.
And so, we again conclude that \hbox{$\alpha_1=\alpha_2$}. This proves that
there is a unique limit point, hence the convergence of the sequence.\qed 

\vskip 5pt
\subsection{Proof of Lemma~\ref{close}}

We can now turn to the proof of Lemma~\ref{close}.

We already know that the sequence is stepwise, that it has no alternation, that it is
not ultimately direct and that it has at least one limit point, say~$\alpha$.
From Lemma~\ref{linepenta}, we may replace~$\ell$ by a
line~$\lambda$ of the pentagrid which does not pass through~$\alpha$.
Let $H_1$ be the half-plane delimited by~$\lambda$ which touches~$\alpha$ and~$H_2$
be its complement in~$\mathbb{H}^2$. It is also plain that if we find a quarter~$F_n$
satisfying the conclusion of the lemma, this will also be the case for all $F_m$'s with
\hbox{$m\geq n$}.

There is a first~$n$ such the vertex~$S_n$ of the quarter~$F_n$ is in the interior of~$H_1$.
Accordingly, the head~$P$ of~$F_n$ has a side on~$\lambda$ and~$S_n$ is either
one of its two vertices at the distance~$a$ from~$\lambda$ or the single one at the distance~$b$.
In the latter case we are done: the rays~$r_1$ and~$r_2$ issued from~$S_n$ have both a
common perpendicular with~$\lambda$ so that $F_n$ contains~$H_2$.  

Now, assume that~$S_n$~is at the distance~$a$ from~$\lambda$. Let $m$ be the first integer
not smaller than~$n$ such that \hbox{$F_m\sqsubset_0 F_{m+1}$}. As there is no alternation,
$S_m$ is on the same line~$\ell_1$ passing through~$S_m$ and~$S_n$ which is perpendicular
to~$\lambda$. As the sequence is not ultimately direct, there is such an~$m$.
From the non-alternation assumption, the head $P_{m+1}$ of~$F_{m+1}$ has one ray~$r_1$ of its
border which is perpendicular to~$\ell_1$ and the other ray $r_2$ is perpendicular to~$r_1$
and it lies in the same side of~$\ell_1$ as~$P_m$. Then $S_{m+1}$ is the vertex which is opposite
to the side~$e$ of~$P_{m+1}$ shared with~$P_m$. Now, $\ell_1$ is a common perpendicular to~$r_1$
and to~$\lambda$, so that $r_1$ lies in~$H_1$. Now, $r_2$ has a common perpendicular with 
the line~$\mu$ which supports~$e$.
Now, $\mu$ itself is perpendicular to~$\ell_1$, so that it is contained in~$H_1$. Accordingly,
$r_2$ is also contained in~$H_1$, as it is on the other side of~$\mu$ with respect to~$\lambda$.
This proves that $F_{m+1}$ contains~$H_2$.\qed 

\section{Two non-computability results}
\label{nocomput}

Theorem~\ref{to_end} 
makes use of the 
compacity theorems which are not algorithmically true. We shall use the tools
used in the proof of Theorem~\ref{to_end} to prove that it is algorithmically
impossible to say whether two given ends are equal or not.

   The theorem says that in a sequence of embedded quarters, their vertices converge to a limit
which is a point at infinity of the hyperbolic plane. We can easily be convinced that
a quarter can be clearly identified by three vertices of its head~$P$: the vertex~$S$ of the 
quarter and the two vertices $A$ and~$B$ of~$P$ which are joined to~$S$ by an edge of~$P$. 
Call \textbf{hat of the quarter} the triple $ASB$ or $BSA$. The rays defining the quarter
are defined by~$SA$ and~$SB$ with $S$ being the point from which the ray is issued and the 
second point being a point on the ray.
As each vertex
can be identified by a coordinate, see for instance~\cite{mmbook1,mm_erzsy}, the hat of a
quarter is a piece
of information which can easily be encoded for an algorithm. 
In an algorithmic approach, a sequence of embedded quarters
is an algorithm, which, in principle, can also finitely be encoded. The embedding condition
can also be encoded, much more easily if we assume the sequence to be stepwise 
with no-alternation. However, the fact that there is no alternation cannot algorithmically be 
checked and the stepwise condition also cannot algorithmically be checked: intuitively,
this would require an infinite time. The algorithmic translation
of Theorem~\ref{to_end} translates the sentence \textit{to each sequence of embedded quarters,
we can define an end to which the sequence of their vertices converge}. This notion 
of convergence means that it is possible to assign to each line~$\lambda$ of the pentagrid a 
rank~$N$ which ensures that the quarters with a higher rank are beyond~$\lambda$ and that
the sequence of the~$\lambda$'s define an end. We may assume that the lines of the pentagrid
can also be encoded, for example, by a pair of vertices of the pentagrid. The convergence of this
sequence of lines to an end cannot be checked but it nonetheless can be defined.
Indeed, from Lemma~\ref{linepenta}, if a line~$\ell$ defines a half-plane containing 
an end~$\alpha$, there is a line of the pentagrid~$\lambda$ which defines a half-plane also 
containing~$\alpha$. This allows us to consider the ends 
which can be defined by a sequence of lines of the pentagrid.

   And so, to each sequence of quarters, we associate a sequence of lines of the pentagrid
which defines the end and this translation from a sequence of quarters to a sequence of lines
of the pentagrid must be algorithmic. Assume also that two lines of the pentagrid being given,
it is possible to decide whether they define non-intersecting half-planes or not.
Now we show that it is not possible to algorithmically distinguish given ends.

\begin{thm}\label{noalgo}
There is no algorithm which would for any sequence of lines of the pentagrid
defining ends whether these ends are equal or not.
\end{thm}

\noindent
Proof. The proof consists in constructing a sequence of sequences of quarters for which
there is no algorithm defining an end.
We define the sequence of sequences as follows. First, we need an algorithmic ingredient:
it is the Kleene function, \hbox{${\cal A}(m,n,k)$} which takes value~1 if the $k^{\rm th}$
step of computation of the $m^{\rm th}$ Turing machine halted on the data encoded by~$n$
and it takes value~0 if this is not the case. Note that if
\hbox{${\cal A}(m,n,k)=1$}, then \hbox{${\cal A}(m,n,k+1)=1$}.

Our algorithm works as follows. Fix a line of the pentagrid, say~$\delta_0$ which passes 
through~$O$, a vertex of the pentagrid which we fixed once and for all. 
Fix~$\eta_0$ the other line of the pentagrid which passes through~$O$. Define~$P_0$ to
be a pentagon with vertex~$O$. Call~$\alpha_0$
the end of~$\delta_0$ which is not in the same side as~$P_0$ with respect to~$\eta_0$.
Define $A_0$ to be the other vertex of~$P_0$ on~$\eta_0$ and $B_0$ to be the other vertex 
of~$P_0$ on~$\delta_0$. The hat of~
$F_0$ is then defined by the triple $A_0OB_0$.
For each~$n$ and~$k$, we define a quarter of $F_{n,k}$ by its head~$P_{n,k}$ and 
its hat: 
$A_{n,k}S_{n,k}B_{n,k}$.
Define 
\hbox{$P_{0,0}=P_0$}, \hbox{$S_{0,0}=O$}, \hbox{$A_{0,0}=A_0$},
\hbox{$B_{0,0}=B_0$} and \hbox{$S_{n,-1}=B_0$}. We define a flag by~\hbox{$f=0$}.
\vskip 5pt
{\leftskip 20pt\parindent 0pt
- As long as \hbox{${\cal A}(n,n,k+1)=f$}, 
$P_{n,k+1}$ is the reflection of~$P_{n,k}$ in~$S_{n,k}A_{n,k}$; 
$S_{n,k+1}$ is the reflection of~$S_{n,k-1}$ in~$S_{n,k}A_{n,k}$ too,
$A_{n,k+1}$ is the other end of the side of~$P_{n,k+1}$ which
passes through~$S_{n,k+1}$ and which is orthogonal to~$\delta_f$, $B_{n,k+1}$ is $S_{n,k}$.
\vskip 0pt
- If \hbox{${\cal A}(n,n,k+1)=1$} and \hbox{$f=0$}, then \hbox{$f:=1$};
$P_{n,k+1}$ is still the reflection of~$P_{n,k}$ in~$S_{n,k}A_{n,k}$; $A_{n,k+1}$ is the vertex
of $P_{n,k+1}$ which is the reflection of~$B_{n,k-1}^\ell$ in~$S_{n,k}A_{n,k}$,
$S_{n,k+1}$ is the other end of the side of~$P_{n,k+1}$ which passes through~$A_{n,k+1}$
and which is orthogonal to~$\delta_0$,
$B_{n,k+1}$ is the other end of the side of~$P_{n,k+1}$ which
ends at~$S_{n,k+1}$ and which does not meet~$\delta_0$, \hbox{$S_{n,k-1}=B_{n,k+1}$},
let $\delta_1$ be the line defined by $S_{n,k+1}B_{n,k+1}$.
\par}

It is clear that for each~$n$, the sequence \hbox{$\{F_{n,k}\}_{k\in\mathbb{N}}$}
is a sequence of embedded quarters. The sequence is stepwise and it has no alternation
by construction.
If the Turing machine numbered by~$n$ does not halt on the data~$n$, then
\hbox{${\cal A}(n,n,k)=0$} for all~$k$ and so, the sequence is ultimately direct, it is even
direct, so that,
by Lemma~\ref{non_alternate}, the sequence $S_{n,k}$ converges to~$\alpha_0$.
If the Turing machine numbered by~$n$ halts on the data~$n$, there is an integer~$m$
such that \hbox{${\cal A}(n,n,m)=0$} and \hbox{${\cal A}(n,n,m+1)=1$}. Accordingly,
\hbox{$F_{n,m}\sqsubset_0 F_{n,m+1}$} but, afterwards, the sequence 
satisfies \hbox{$F_{n,k}\sqsubseteq_0 F_{n,k+1}$} and \hbox{$F_{n,k}\not\sqsubset_0 F_{n,k+1}$}.
The sequence is again ultimately alternate but, this time, it converges to the end $\alpha_1$ 
of~$\delta_1$ which is contained in the other side of~$\eta_0$ with respect to~$P_0$.
Now, By construction, as $S_{n,m+1}A_{n,m+1}$ is a common perpendicular to~$\delta_0$
and~$\delta_1$, these lines are non-secant, in particular, they cannot be parallel. 
Accordingly, \hbox{$\alpha_0\not=\alpha_1$}.

Now, if \hbox{$\alpha_0\not=\alpha_1$}, among the lines of the pentagrid
which defines these ends, we can find two of them $\lambda_1$ and~$\lambda_2$ such that
denoting by $\pi_1$, $_pi_2$ the half-plane defined by $\lambda_1$, $\lambda_2$
respectively and which touches $\alpha_1$, $\alpha_2$ respectively, 
we get \hbox{$\pi_1\cap\pi_2=\emptyset$}. And this can be performed algorithmically
if \hbox{$\alpha_0\not=\alpha_1$}.
Now, if we had an algorithm which could tell us whether these limits are the same
or not, this algorithm could be used to decide the halting problem for Turing machines,
which is known to be impossible.\qed

Now, we can prove another result of the same flavor.

\begin{thm}\label{noconv}
We can construct a sequence \hbox{$\{G_{k,n}\}_{k\in\mathbb{N}}$} of quarters
whose vertices are $S_{k,n}$ such that for each~$n$ the
sequence \hbox{$\{S_{k,n}\}_{k\in\mathbb{N}}$} converges to a point at infinity~$y_n$ 
and such that the sequence \hbox{$\{y_n\}_{n\in\mathbb{N}}$} cannot algorithmically
converge to any point at infinity.
\end{thm}

Note that if the sequence \hbox{$\{y_n\}_{n\in\mathbb{N}}$} converges, it must
converge to a point at infinity.

\noindent
Proof of Theorem~\ref{noconv}.
Consider the same function \hbox{${\cal A}(n,n,m)$} as previously. We change the construction
as follows. We construct a sequence of sequences
\hbox{$\{G_{m,n}\}_{m\in\mathbb{N}}$}, again defining a quarter
by its head and its hat. In what follows, the hat will be given as previously but
the order of the vertices is important. In a tile $P_{k,n}$, we consider that the
hat is~$A_{k,n}S_{k,n}B_{k,n}$. Let~$e_{k,n}$ be the side which is opposite 
to~$S_{k,n}$. We consider that $e_{k,n}$~is
at the \textbf{bottom} of the tile, that~$S_{k,n}$ is at its top, that $A_{k,n}$~is at 
its left-hand side and~$B_{k,n}$ at its right hand-side: we can consider that starting 
from~$A_{k,n}$ and clockwise
turning around the tile we meet~$S_{k,n}$, then~$B_{k,n}$ and then the ends of~$e_{n,k}$. 
The side~$A_{k,n}S_{k,n}$ will be called side~0 and the side~$S_{k,n}B_{k,n}$ will be called 
side~1. 
For each fixed~$n$, we start with~a fixed once and for
all tile with bottom~$e_0$ and hat~$A_0S_0B_0$ which will be denoted by~$P_{0,n}$.
At the beginning \hbox{$k=0$}. Then we construct the sequence as follows.

\vskip 5pt
{\leftskip 20pt\parindent 0pt
- If \hbox{${\cal A}(k+1,k+1,n)=0$}, 
$P_{k+1,n}$ is the reflection of~$P_{k,n}$ in~$A_{k,n}S_{k,n}$; $e_{k+1,n}$ is 
$A_{k+1,n}S_{k+1,n}$, which fixes the hat with the conventions we have already defined.
We say that $P_{k+1,n}$ has the value~0.
\vskip 0pt
- If \hbox{${\cal A}(k+1,k+1,n)=1$}, then 
$P_{k+1,n}$ is the reflection of~$P_{k,n}$ in~$S_{k,n}B_{k,n}$; $e_{k+1,n}$ is 
$S_{k,n}B_{k,n}$.
We say that $P_{k+1,n}$ has the value~1.
\par}

$G_{k,n}$ is the quarter defined by~$P_{k,n}$ and its hat $A_{k,n}S_{k,n}B_{k,n}$.
By construction, it is plaint that for each $n$ and~$k$ we have
\hbox{$G_{k,n}\sqsubset_0 G_{k+1,n}$}. Accordingly, the sequence
\hbox{$G_{k,n}\sqsubset_0 G_{k+1,n}$} is stepwise, with no alternation and it is not
ultimately direct: when $n$ is fixed, there is always a Turing machine numbered with~$k$
such that its computation on~$k$ is completed at the $n^{\rm th}$ step. 
From Theorem~\ref{to_end}, the sequence 
\hbox{$\{S_{k,n}\}_{k\in\mathbb{N}}$} tends to a point at infinity~$y_n$.
By construction of the quarters, we can notice that 
for each~$k$ we the rest of the
sequence evolves in \hbox{$\mathbb{H}^2\backslash G_{k,n}$}. Fix~$k$
and let \hbox{$K_0=\mathbb{H}^2\backslash G_{k,n}$} when \hbox{${\cal A}(k,k,n)=0$}
and \hbox{$K_1=\mathbb{H}^2\backslash G_{k,n}$} when \hbox{${\cal A}(k,k,n)=1$}.
It is not difficult to see that \hbox{$K_0\cap K_1=\emptyset$}. More than that,
the line defined by $S_{k,n}B_{k,n}$ for~$K_0$ and the line defined by
$A_{k,n}S_{k,n}$ for~$K_1$ are non-secant. This means that the distance between~$K_0$
and~$K_1$ tends to infinity when we go to infinity on both these borders.
From this remark, assume that the sequence \hbox{$\{y_n\}_{n\in\mathbb{N}}$}
tends to a limit~$y$ which is also a point at infinity. Then there is a half-plane~$H$
delimited by a line~$\lambda$ which may be assumed to belong to the pentagrid such
that there is $N$ such that for~\hbox{$n\geq N$}, all $y_n$'s are touched by~$H$.
By the remark we made about~$K_0$ and~$K_1$, we can see that, necessarily,
for any \hbox{$n\geq N$}, $G_{k,n}=G_{k,N}$, otherwise, $y_n$ and~$y_N$ cannot be both in~$H$.
Accordingly, if we have an algorithm~$\varphi$ which, for each~$n$, gives an integer~$\varphi(n)$
such that all $y_p$ with \hbox{$p\geq \varphi(n)$} are in~$H$ which is at distance~$n$ from~$P_0$,
then, looking at the value of $G_{n,\varphi(n)}$, we know whether
\hbox{${\cal A}(k,k,n)=0$} for ever or not. And this decides the halting problem, 
which is impossible. \qed

\section{conclusion}

   Probably, other undecidability results of analysis can be transported into the hyperbolic
plane in  similar way. This might open a new area.

\end{document}